# Wheel-Rail Interface Condition Estimation (W-RICE)

Sundar Shrestha[1, 2, *], Anand Koirala[3], Maksym Spiryagin[1], Qing Wu[1]

[1]Centre for Railway Engineering, Central Queensland University, Rockhampton, Queensland, Australia

[2]Rail Manufacturing Cooperative Research Centre, Melbourne, Victoria, Australia

[3]School of Health, Medical and Applied Sciences, Central Queensland University, Rockhampton, Australia

## Abstract

The surface roughness between the wheel and rail has a huge influence on rolling noise level. The presence of the third body such as frost or grease at wheel-rail interface contributes towards change in adhesion coefficient resulting in the generation of noise at various levels. Therefore, it is possible to estimate adhesion conditions between the wheel and rail from the analysis of noise patterns originating from wheel-rail interaction. In this study, a new approach to estimate adhesion condition is proposed which takes rolling noise as input.

Acoustic sensors (Behringer B-5 condenser microphone) for audio data acquisition were installed on a scaled bogie test rig. The cardioid configuration of the sensor was chosen for picking up the source signal while avoiding the surround sound. The test rig was operated at the speed of 40 and 60 rpm multiple times in both dry and wet friction conditions. The proposition behind running the setup several times was to get a large set of acoustic signals under different adhesion conditions because there exists no such dataset in the public repositories. 30 seconds interval of rolling noise data from the continuous audio signal was extracted as samples for training W-RICE model. Each sample was pre-processed using Librosa python module to extract seven basic features/signatures: zero-crossing rate, spectral centroid, spectral bandwidth, spectral roll-off, MFCCs (Mel-frequency cepstral coefficients), RMS (root-mean-square) energy and Chroma frequencies. These features were used as input to an MLP (Multi-Layered Perceptron) neural network and trained for four different classes (dry and wet conditions with the speed of 40 rpm and 60 rpm each). MLP model training and testing were implemented in Keras deep learning library. MLPs can automatically learn useful and meaningful relationships among the input features for classifying the input audio sample to one of the classes that were trained for. For four classes considered, 100% classification accuracy was achieved on the test set while accuracies of 99.56%, 73.24%, and 63.59% were achieved for three validation sets consisting of audio samples of varying noise levels.

*Corresponding author: s.shrestha@cqu.edu.au

# 1. Introduction

There are different types of noise in railway such as rolling noise, impact noise, curve sequel, bridge noise, aerodynamic noise, ground noise, and others (shunting noise, warning noise, and traction noise) as mention in the first chapter of [1]. Those noises are generated from the influence of different parameters such as the roughness of contact interface, number of wheels, wheel-rail defects, wheel load on the contact patch, rail vehicle speed, wheel-rail geometry, etc. [2]. Since this research domain is focused on the wheel-rail interface, only the noise emitted from the interaction between wheel and rail will be considered in this paper. In such scenarios, three types of noise can be analyzed. They are rolling noise, impact noise, and curve sequel. The rolling noise is generated due to the vibration of wheel and rail excited by surface ripples (roughness) and propagated into the surrounding in the form of sound. The impact noise is generated when larger discrete features such as rail joint, wheel flat, dipped welds and gaps at points and crossing occurs. The squealing noise is generated when the rail vehicle travels around tight curve tracks [3–7].
Most of the wheel-rail noise analysis has been conducted to mitigate those noise. The noise analysis has been performed to reduce rolling noise in reference [8] and to reduce curve squeal in reference [6,7]. In reference [9], prediction of the angle of attack between the wheel and rail-based on noise analysis has been performed. A series of experiments were made with a test rig and the obtained data was recorded for different values of the angle of attack and studied by analyzing the generated spectrograms for each sound. Furthermore, the wheel-rail noise was analyzed to determine the effects of friction modifiers [10] and to monitor rail [11]. In real-world, the noise emitted in the same condition may slightly differ at different times. Thus, even a single condition can hold several slightly different noise signals. However, in all of the above studies, the data to analyze noise was limited which make the friction condition estimation process less accurate by not recognizing a slight difference noise signal.

Moreover, the literature [12–14] suggest that the roughness present between wheel and rail has a huge influence on rolling noise level. That means the presence of third body layer and its contribution to change friction coefficient has also a big influence on noise generation. Thus, it is possible to develop an adhesion condition detection system depending on noise analysis in the wheel-rail contact. Since the roughness has a greater influence on rolling noise as compare to impact noise and curve squeal, only the rolling noise will be considered in this paper. The frequency range of the rolling noise is between 50 Hz to 5000 Hz, in which, rail radiated noise lies within 0.5…1 kHz and wheel radiated noise lies above 1 kHz [1].

# 2. Methodology

The estimation task in this project is divided into data acquisition, model preparation, model validation, and sound event detection. In this project, the sound event refers to a sound produced by wheel-rail interaction. Since there are no such public repositories for such sound events, a huge amount of data is collected and managed. This is followed by learning model preparation and validation as shown in Figure 1-(A). The validated model is further implemented to detect and classify the sound event to estimate the friction condition as shown in Figure 1-(B).

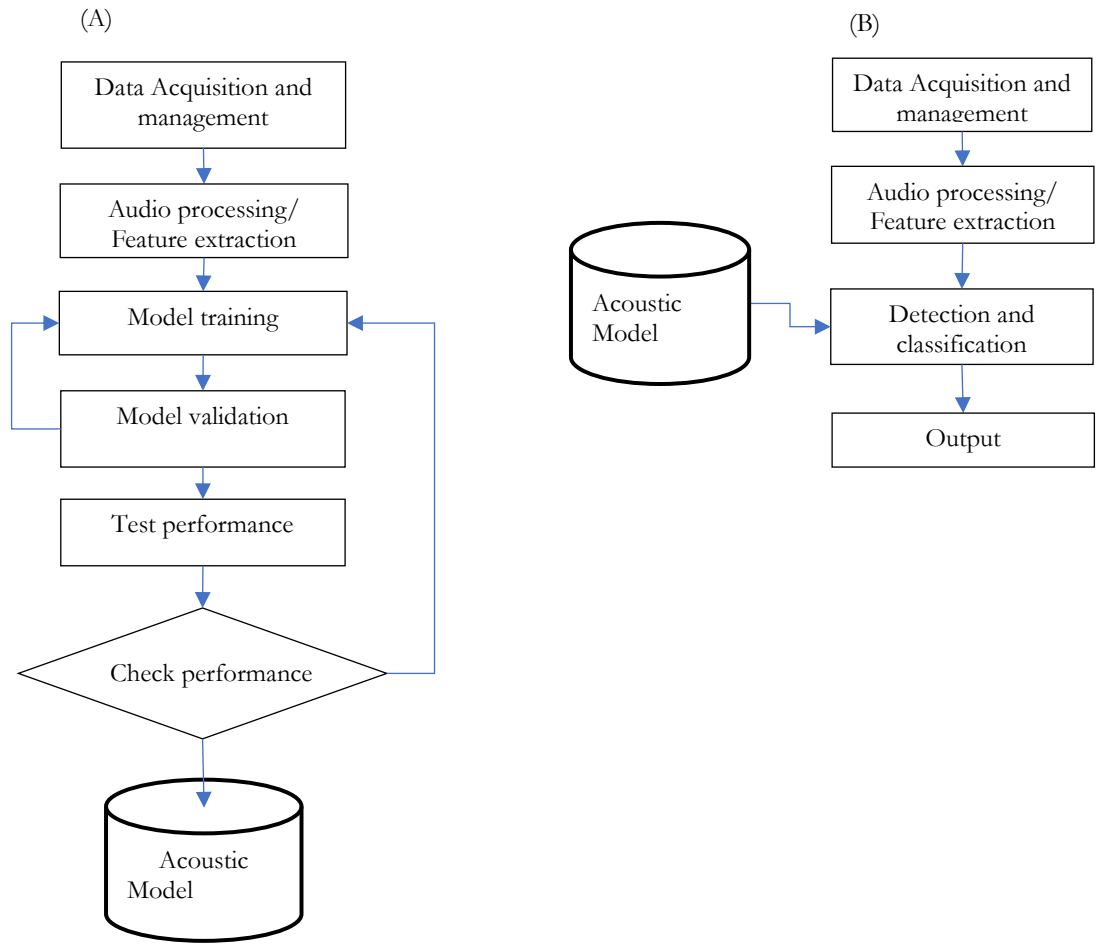

Figure 1: (A) Methodology to prepare a learn model, (B) Audio analysis system structure

## 3. Data Acquisition and Management

The acoustic datasets are important to learn the parameter of the acoustic model as well as evaluate the performance of the learned model. Thus, data acquisition is very crucial since the accuracy of the system depends highly on the variety of data used. To get such dataset, acoustic signals in all possible conditions intended to target applications need to be recorded/stored/collected. In simple word, enough representative example of all sound classes helps to develop a better model. The metadata should contain true information of the target application, which are annotated manually during data collection.

The acoustic signal may vary due to the acoustic environment, capture device, relative placement and interfering noise source. For optimal performance, the above factors should be consistent between training and actual usage stage. Since some of the factors cannot be controlled which leads to poor performance. Thus, mixed-condition training is a better approach in which the training materials (i.e. acoustic signal) under different conditions are provided in learning stage. Most of the time reference annotations are obtained manually which is time-consuming and sometimes difficult if the signal is too noisy.

One of the key challenges in this project is to install an acoustic sensor that is sensitive to the rolling noise and reasonably robust to withstand the operational environment. There are other concurrent noises originating from different parts of the test rig arrangements. To avoid such unwanted noise and to capture the rolling noise, a sensor (Behringer B-5) is installed close to wheel-roller interface as shown in Figure 2. The adjustment of the sensor is perpendicular to the direction of the wheel rotation. It has two interchangeable capsules: omnidirectional for picking up sound on all sides of the microphone, and cardioid for picking up the source signal while avoiding the axis sound. The latter capsule with the low-cut filter option is used in this project to eliminate infrasonic caused by surround sound and turbulence. An audio USB interface (Behringer U-Phoria UMC204HD) having 24-Bit/192 kHz resolution has been used for acoustic signal acquisition. The test set up is run at 40 rpm and 60 rpm multiple times in dry and wet wheel-roller friction conditions. The proposition behind running the setup several times is to get a large set of acoustic signals under different friction conditions. Currently, there exists no such dataset in the public repositories. Thus, collection of abundant data ensures the sensitivity, robustness, and accuracy of the analysis and classification techniques.

The abundantly collected data are further arranged for easy accessibility. This requires, converting data to .wav format, arranging the dataset with associated metadata (speed, dry friction, wet friction), preparing datasets for training and validation purposes. The considered categories in this project are 'wet friction condition' and 'dry friction condition'. Figure 3 shows the example result obtained from noise analysis representation of data segment relating to these two categories.

## 4. Audio processing and Feature extraction

The purpose of the feature extraction is to get adequate information for detection and classification of the target sound which makes the modelling phase easier and computationally cheaper. In the development phase, the learn acoustic model is prepared by using the reference annotations together with the acoustic features from audio processing. Audio processing is performed to transform the audio signals into a compact representation suitable for machine learning. Two phases are there: pre-processing in which the noise effect of audio signal is reduced to highlight the target sound, feature extraction in which audio signal is changed into compact representation. The pre-processing is used as required only. For recognition, it is better to have less variation among the feature extracted from the materials allocated to same category and simultaneously, high variation between extracted features from different categories. The compact feature required less memory and computational power to process as compared to the original audio signal.

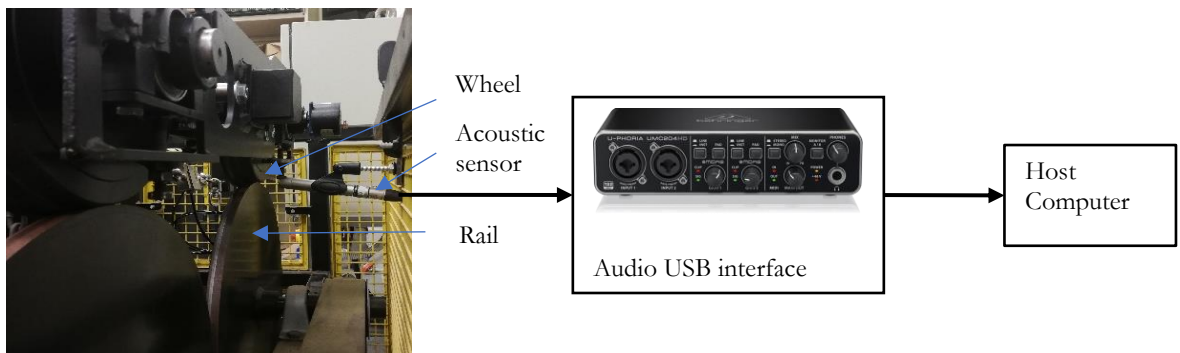

Figure 2. Installation of the acoustic sensor on test rig and setup for noise data acquisition.

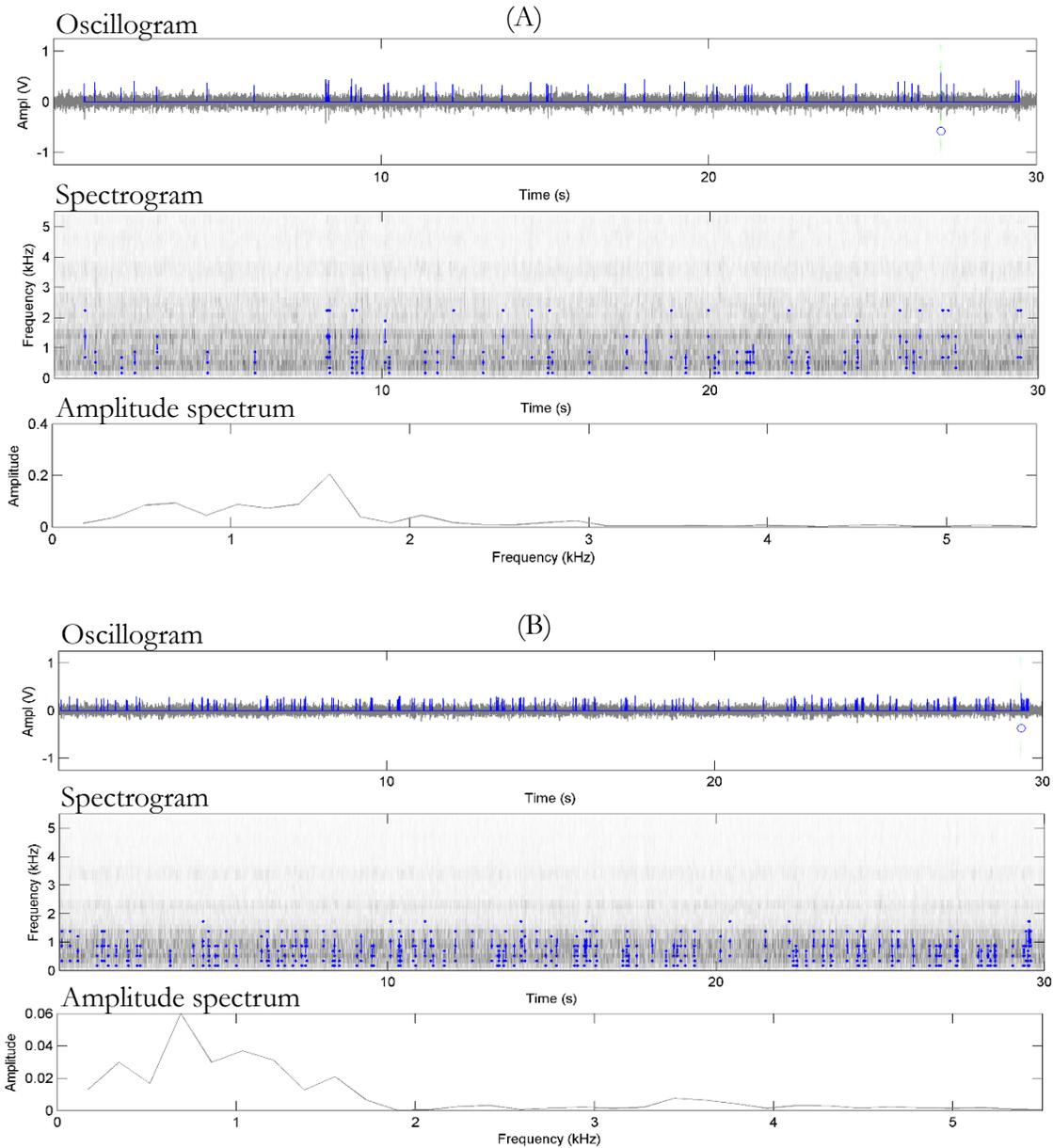

Figure 3. Example results from noise analysis at each event type: dry friction condition (A) and wet friction condition (B) at the speed of 60 rpm

A spectrogram is a visual representation of the frequency spectrum of a signal as it varies with time. Deep learning convolution neural network (CNN) models have been extensively used for image classification and object recognition tasks. However, there are difficulties in using spectrogram as input to the CNN models for classification. Unlike object detection, where an arbitrary set of neighboring pixels belong to a particular object category, CNN cannot differentiate whether a specific frequency at a certain time originated from single or multiple sources because the sound frequencies are most often non-locally distributed on spectrogram. Similarly, object feature extracted from images using CNN carries the same meaning regardless of their position from image axes while spectrogram axes (frequency and time) carries different meaning. Moreover, sound is inherently serial and occurs as events rather than objects in images which are more often visually static in the scenes. Therefore, it is difficult for CNNs to analyze and extract meaningful information directly from spectrograms for audio classification. However, some features can be extracted from the audio signal and fed into neural networks for finding meaningful relationships among the input features to accurately classify audio samples. In the current study, 7 basic features were extracted from the audio samples at different conditions of the wheel-rail interface to train an MLP (Multi-Layered Perceptron) neural network for audio classification for friction condition estimation.

Some of those features extracted from the wheel-rail interface noise in this project are:
- Zero crossing rate:

The rate of changing the sign of the audio signal from positive to negative or vice versa.
- Spectral centroid

It is the center of mass of the sound calculated as the weighted mean of the frequencies in the sound.
- Spectral roll-off

It measures the shape of the signal. It characterizes the frequency lower than a specified percentage of the total spectral energy.
- Mel-frequency cepstral coefficients

The Mel frequency cepstral coefficients (MFCCs) of a signal are a small set of features (20 in this study) which briefly describe the overall shape of a spectral envelope.
- Chroma frequencies

Chroma features are an interesting and powerful representation for music audio in which the entire spectrum is projected onto 12 bins representing the 12 distinct semitones (or chroma) of the acoustic octave.
- Spectral bandwidth

This computes the $2^{nd}$ order spectral bandwidth by default whose order can be changed.
- Root-mean-square energy

It computes root-mean-square (RMS) energy for each frame from the audio samples. Computing the energy from audio samples is faster as it does not require a short-time Fourier transform calculation.

## 5. Model training and validation

The audio datasets have categories dry_40, dry_60, wet_40 and wet_60 having 52, 61, 51 and 64 audio samples respectively. For each sample in the 4 categories considered 'librosa' python package was used to extract 7 basic audio features: - zero-crossing rate, spectral centroid, spectral roll-off, MFCC (Mel-frequency cepstral coefficient), chroma frequencies, spectral bandwidth, and root-mean-square energy. 20 MFCCs were extracted resulting in total 26 input features for training.

For model training and testing samples from each category were shuffled and split into train_set (80%) and test_set (20%) using 'train_test_split' function from Sklearn. The dataset was standardized using 'StandardScaler' from Sklearn for MLP (Multi-Layered Perceptron) model training and testing. Similarly, labels were encoded to numbers between 0 and 3 for the 4 categories considered using 'LabelEncoder' from Sklearn for MLP training.

The MLP consisted of 4 layers: - 1 input layer, 2 hidden layers and 1 output layer. The input and hidden layers consisted of 512 neurons each and 'relu' activation function. The output layer consisted of 4 neurons for 4 output categories/classes and 'softmax' activation function. MLP model was compiled with 'adam' optimizer (learning rate =0.01) and 'sparse_categorical_crossentropy' loss function. The MLP model was trained on train_set for 60 epochs with batch size of 32.

## 6. Result

The trained model was saved and used for category label prediction on the test set. On the test_set a 100% classification accuracy was achieved. For further validation of the model, 3 other datasets were created by augmenting the original dataset with several levels of random noise using 'randn' function from 'numpy' package. Generated random floats were multiplied by 0.5, 0.05 and 0.005 and added to the original audio files to create valid_0.5, valid_0.05, and valid_0.005 datasets. Classification accuracies of 99.56%, 73.24% and 63.59% was achieved for validation sets valid_0.005, valid_0.05 and valid_0.5 respectively.

## 7. Discussion and conclusion

In the current study audio samples of 30 seconds, each was used but it could be possible to achieve similar results with smaller samples. To validate the model several levels of random noise were added to the dataset achieved from the test rig. The model was able to tolerate small level of noise but it would have been better if the training dataset consisted of more variations (different noise levels). Thus, it would be better if the audio from real field could be used. Further, more experiment could be done with the neural network to find out the most important features that are sufficient for classification. Small samples and less features help to reduce the computational competency and will be quicker for real-time implementation.

## Acknowledgments

The authors greatly appreciate the financial support from the Rail Manufacturing Cooperative Research Centre (funded jointly by participating rail organisations and the Australian Federal Government's Business Cooperative Research Centres Program) through Project R1.7.1 – "Estimation of adhesion conditions between wheels and rails for the development of advanced braking control systems".